\documentclass[useAMS,usenatbib]{mn2e}

\usepackage{txfonts}
\usepackage[dvips]{graphicx}
\usepackage{epsfig}
\usepackage{natbib}
\usepackage{subfigure}
\usepackage{multirow}
\bibpunct{(}{)}{;}{a}{}{,}

\newcommand{\kms}{{\rm km~s^{-1}}}
\newcommand{\sm}{{\rm M_{\odot}}}

\title[HI mass and velocity functions]%
{The low mass end of the neutral gas mass and velocity width functions of galaxies in 
  $\Lambda$CDM }

\author[Yaryura et al.]
  {C.Y.~Yaryura$^{1}$, A.~Helmi$^{2}$, M.G.~Abadi$^{1}$and E.~ Starkenburg$^{3}$\\ 
 $^{1}$Instituto de Astronom\'{\i}a Te\'{o}rica y Experimental
  (CONICET-UNC). Observatorio Astron\'{o}mico de C\'{o}rdoba, 
  Laprida 854, X5000BGR, C\'{o}rdoba, Argentina\\
 $^{2}$ Kapteyn Astronomical Institute, University of Groningen, P.O. Box 800, 9700 AV Groningen, The Netherlands\\
$^{3}$Leibniz-Institut f\"ur Astrophysik Potsdam (AIP), An der Sternwarte 16, D - 14482, Potsdam, Germany}

\date{}
\pagerange{\pageref{firstpage}--\pageref{lastpage}} \pubyear{}

\def\LaTeX{L\kern-.36em\raise.3ex\hbox{a}\kern-.15em
    T\kern-.1667em\lower.7ex\hbox{E}\kern-.125emX}

\begin{document}
\label{firstpage}
\maketitle

\begin{abstract}
  We use the high--resolution Aquarius cosmological dark matter simulations coupled to the
  semi--analytic model by \citet{Starkenburg:2013} to study the HI content and
  velocity width properties of field galaxies at the low mass end in the context of $\Lambda$CDM. We
  compare our predictions to the observed ALFALFA survey HI mass and
  velocity width functions, and find very good agreement without
  fine--tuning, when considering central galaxies. Furthermore, the properties of the dark matter halos
  hosting galaxies, characterised by their peak velocity and circular
  velocity at 2 radial disk scalelengths overlap perfectly with the
  inferred values from observations. This suggests that our galaxies
  are placed in the right dark matter halos, and consequently at face value, we do
  not find any discrepancy with the predictions from the
  \mbox{$\Lambda$CDM} model. Our analysis indicates that previous
  tensions, apparent when using abundance
  matching models, arise because this technique cannot be
  straightforwardly applied for objects with masses $M_{vir} < 10^{10}
  \sm$.
\end{abstract}

\begin{keywords}
galaxies: luminosity function, mass function -- galaxies: kinematics and dynamics -- galaxies: dwarf -- galaxies: halos
\end{keywords}


\section{Introduction} \label{S_intro}

The observed properties of our universe on large--scales are very well
reproduced by the current standard $\Lambda$CDM cosmological model
\citep{Komatsu:2011}. However, on small scales the paradigm faces a
number of challenges that could be related to inherent model flaws or
due to our poor understanding and modelling of the physical processes
that affect galaxies on these scales.

A well known problem is the excess of low--mass dark matter halos
compared with the actual number of visible low--mass galaxies.  This
discrepancy is commonly known as the CDM overabundance problem or
missing satellite problem \citep{Moore1999,Klypin1999}. One
manifestation of this is evident in the faint--end of neutral hydrogen
(HI) mass functions \citep{Papastergis:2011} and in galaxy luminosity
functions \citep{Klypin:2014}. For example, dark matter only
simulations predict a slope for the mass--function of $\alpha \sim
-1.8$, which is in contrast to the significantly shallower slope
$\alpha \sim -1.3$ exhibited for example by the HI mass function of
galaxies in the ALFALFA survey \citep{Martin:2010}.

The mass and luminosity functions relate to properties of galaxies,
but velocity functions are perhaps more revealing because they are
connected to the internal properties of the host dark matter
halos. \citet{Papastergis:2014} and \citet{Klypin:2014} have recently
drawn attention to the fact that the velocity function of dark matter
halos $dN/d\log V \sim V^{\alpha}$ has $\alpha \sim -3$, while
observationally $\alpha \sim -1$ for galaxies with circular velocities
smaller than $\sim 60~\kms$.

Furthermore, it has been pointed out that CDM halos hosting
low mass galaxies would be too concentrated in comparison to what can
be inferred from observations, such as from rotation curves
\citep{Ferrero:2012} or velocity dispersions at the half--light radii of
dwarf galaxies \citep{Boylan2012}. This is known as the ``too
big to fail problem'' which has recently drawn a lot of attention in the literature.

We can classify the solutions to these problems in two types: those
related to baryonic physics and those that propose changes in the
nature of dark matter \citep[see e.g.][ for a
review]{Kravtsov2010}. Among the many baryonic effects considered we
can mention photoheating during the reionization epoch
\citep[e.g.][]{Barkana:1999, Bullock:2001a, Shapiro:2004} which results in a
lower baryon fraction in low mass halos, the inability (or
inefficiency) of HI cooling for halos with virial temperature below
10$^4$K \citep{Haiman2000}, and stellar feedback \citep[e.g.][]{Dekel:1986,
  MacLow:1999}. The latter not only acts to lower the baryon fraction
in low mass galaxies, possibly preventing further star formation in
the lowest mass systems, but may also modify the density profiles of
dark matter halos, making them less concentrated
\citep{Governato2012}. On the other hand, warm dark matter or
self--interacting dark matter models, which effectively result in a
suppression in the spectrum of fluctuations of the power on small
scales, predict an important reduction in the number of small dark
matter halos, and possibly systems with lower central densities
\citep[e.g.][]{Colin:2000, Kamionkowski:2000, Bode:2001, Kennedy:2014,
  Polisensky:2014,Vogelsberger2014}
 
Fully cosmological numerical hydrodynamical simulations are just
beginning to model reliably these low mass scales, because of the
difficulty in addressing simultaneously a large volume with high
spatial resolution. Amongst the recent works that have discussed the
issues highlighted above is \cite{Sawala2014} who have found that a
combination of effects related to reionization and environment
introduces strong biases in the halos that host dwarf galaxies, and those
that remain dark.

An alternative approach is to use semi--analytic (SA) models combined for example with
a very high--resolution cosmological N--body simulation. For galaxies
with neutral hydrogen mass $M_{HI} \ge 10^8~\sm$, \citet{Obreschkow2013} show good agreement
between the properties of gas--rich galaxies in the HIPASS survey to
the predictions from a SA model coupled to the
Millenium dark matter simulation \citep{Springel2005}.  For lower masses, and using a
different technique, namely abundance matching,
\cite{Papastergis:2014} argue that the ALFALFA survey should be observing a much
larger number density of dwarf galaxies given the measured
rotational velocities.  Along similar lines,
\citet{Ga-Kimmel2014a} predict there should be of the order of 1000
galaxies with stellar mass $M_* > 10^3 \sm$ to be discovered within $\sim 3$ Mpc
from the Milky Way (MW). Furthermore, assuming a one--to--one relation
between $M_*$ and $M_{HI}$ these authors predict 50 undiscovered
gas--rich dwarf galaxies with $M_{HI} >10^5 \sm$ within the local
volume. However, care must be taken when applying abundance matching at the low mass
end (i.e. virial mass $M_{vir} < 10^{10} \sm$), because this is a regime of strong
stochasticity where galaxy formation may or may not take place in
halos of similar present--day mass depending on whether they were above
or below a given threshold (e.g. for HI cooling) at higher redshift
\citep[see Fig.~16 of][]{Li:2010}.  As we shall see in this Paper
\citep[also in][]{Sawala2014}, for virial masses $M_{vir} \sim 10^{9.5}~\sm$
only $\sim 50\%$ of the halos are expected to host a galaxy. A simple
abundance matching like--technique that only ranks halos by mass around this mass scale will
fail \citep{Sawala:2015}.

\cite{Starkenburg:2013} studied the properties of satellite galaxies
around Milky Way--like halos combining a SA model with the Aquarius
suite of cosmological simulations \citep{Springel:2008a}.  This model reproduced well the
luminosity functions as well as e.g. star formation histories of these
systems, and it was found that the ``too big to fail'' problem could
be solved by invoking a lower total mass for the Milky Way of $\sim 8 \times
10^{11} \sm$ \citep{Vera-Ciro2013}. However, the model satellites
had too high HI fractions, which was attributed to the lack of
ram--pressure stripping of cold gas in the model once a galaxy becomes
a satellite \citep[see e.g.  Fig.~11 of][who use the same SA model but applied to a 
different cosmological simulation]{Li:2010}.

Motivated by the recent HI surveys such as HIPASS and ALFALFA that
probe the lowest mass ends, and by the relative success in solving a
number of problems for low mass satellites, here we use the
\citet{Starkenburg:2013} model to focus on the faint galaxies in the
field. Although the Aquarius simulation high--resolution box is small 
\mbox{($\sim 2.4~h^{-1}$Mpc on a side)}, it is large
enough to contain several hundred small galaxies whose properties can
be contrasted to observations. By focusing on systems in the field, and specifically on central galaxies, we
should also be able to establish if the gas content of our galaxies is
modelled properly and whether the mismatch found for the satellites is
only due to environmental effects.

This paper is organised as follows.  We describe the most relevant
characteristics of our SA method in Section \ref{S_sample}.  In
Section \ref{S_analysis} we compare the luminosity function (LF), HI mass function and the
velocity function for the galaxies in the SA model to those in the
ALFALFA sample from \cite{Papastergis:2014}. We explore in this section
the reasons behind the success of this comparison, and on the failure
of abundance matching methods on the lower mass end.  In Section
\ref{S_Conclusions} we summarise our results and conclusions.


%
\section{Methodology} \label{S_sample}

\cite{Starkenburg:2013} have used the Aquarius dark matter simulations
in combination with a semi--analytic galaxy formation model that stems
originally from \cite{DeLucia:2004,Croton:2006,Delucia2007,Delucia2008,Li:2010}. 
The Aquarius halos were identified in the Millennium--II Simulation
\citep{Boylan-Kolchin:2009}, a cosmological N--body simulation with the
following parameters: \mbox{$\Omega_{m}$ = 0.25},
\mbox{$\Omega_{\Lambda}$ = 0.75}, \mbox{$\sigma_{8}$ = 0.9},
\mbox{$n_s$ = 1}, \mbox{$h = 0.73$} and \mbox{$H_0 = 100 h
  \kms$Mpc}. A series of five zoom--in simulations with progressively
higher resolution centered around 6 different Milky Way mass halos 
were performed, until a particle mass resolution of $m_p \sim 1.7 \times
10^3~\sm$ and spatial resolution of $\sim 20$~pc were achieved
\citep{Springel:2008a}.  
The SA model of \cite{Starkenburg:2013} follows a number of important
physical processes that affect the evolution of a galaxy, including
star formation, feedback, cooling, heating, mergers, etc. We briefly
describe here in more depth those processes that have an important effect on
low--mass galaxies, and refer the reader to the papers mentioned
earlier for more details. In summary: 
\begin{itemize}
\item The {\it feedback} model corresponds to the \citet{DeLucia:2004}
  prescription, in which the gas mass that is reheated by supernovae feedback
  is $\propto E_{SN}/V^2_{vir} \Delta M_*$, i.e. it is inversely
  proportional to the depth of the potential--well (as given by its virial velocity
  $V_{vir}$) and proportional to the amount of newly formed stars
  $\Delta M_*$.

\item \textit{Reionization} is modelled following the simulations of
  \cite{Gnedin:2000}, who quantified the effect of
  photoionization/photoevaporation on low--mass haloes. This
  effectively leads to a reduction in the baryon content in halos
  below a ``filtering mass'',  given by 
\begin{equation}
f_{b, halo} (z, M_{vir}) =
  f_b [1 + 0.26 M_F(z ) /M_{vir}]^3,
\end{equation} 
where $f_b = 0.17$ is the universal baryon fraction. For $M_F( z )$
the analytical fitting function from Appendix B in
\citet{Kravtsov2004} is used\footnote{The 2nd line in Eq.(B2) of this
  paper contains a type--setting error, which has been fixed in the
  version on the ArXiV.}. Reionization is assumed to start at redshift
$z_0 = 15$ and end at $z_r = 11.5$. Although this may be on the
  high redshift end of plausible values \citep{Planck2014}, it is important to realize
  that the Aquarius simulations represent an overdense environment in
  which reionization may well have started earlier.

\item \textit{Cooling} depends on metallicity and temperature of the
  hot gas. Cooling via molecular hydrogen is assumed to be highly
  inefficient and prevented by photo--dissociation caused by UV
  radiation from the stars, especially at early times
  \citep{Haiman2000}, and in the model it is forbidden for halos below
  the atomic hydrogen cooling limit, $T_{vir} = 10^4$ K, where
\begin{equation}
T_{vir} = 35.9 (V_{vir}/\kms)^2.
\end{equation}

\item \textit{Star Formation} in the quiescent mode (as opposed to in
  starbursts) is assumed to take place in an exponential thin disc of radial scalelength $r_d$.
The mass in cold gas of the disc that is in excess
  of critical threshold $M_{crit}$ is transformed into stars:
\begin{equation}
 M_{crit} = 1.14 \times 10^8 (V_{vir}/ 20~\kms) (r_{d}/1~{\rm kpc})
  \sm,
\label{eq:mcrit}
\end{equation}
where this criterion effectively stems from the surface density
critical threshold for star formation found by
\citet{Kennicutt1998}. The value of $r_d$ is calculated as in
  \citet{Delucia2008}, assuming conservation of specific angular
  momentum of the gas as it cools and settles into a rotationally
  supported disk \citep[following][]{Mo:1998}. However, it is recomputed at
  each time-step by taking the mass-weighted average profile of
  the gas disk already present and that of the new material being accreted.

\end{itemize}

The luminosities of our galaxies are computed from the stellar masses
using stellar population synthesis models from \citet{BC:2003}, and 
assuming a Chabrier IMF \citep{Chabrier:2003}, as in \citet{Delucia2007}.

In what follows we use the Aquarius series of halos resolution level--2
(hereafter, Aq-A-2, Aq-B-2, Aq-C-2, Aq-D-2, Aq-E-2 and Aq-F-2) coupled
to our SA model to analyse different galaxy properties and compare
them with observations.  To assess the numerical convergence of the
model we use four resolution levels of the Aq-A halo. Each of these
simulations encompass a high--resolution region of radius $\sim
1.2$~Mpc~$h^{-1}$ that extends well outside the virial radius of the
main Milky Way--like halos.  The ``field (or central) halos'' (i.e. not satellites) located in 
this region are largely the focus of the current study.

\begin{figure}
 \begin{minipage}[t]{\linewidth}\vspace{0pt}\hspace{-25pt}
  \includegraphics[width=1.2\linewidth]{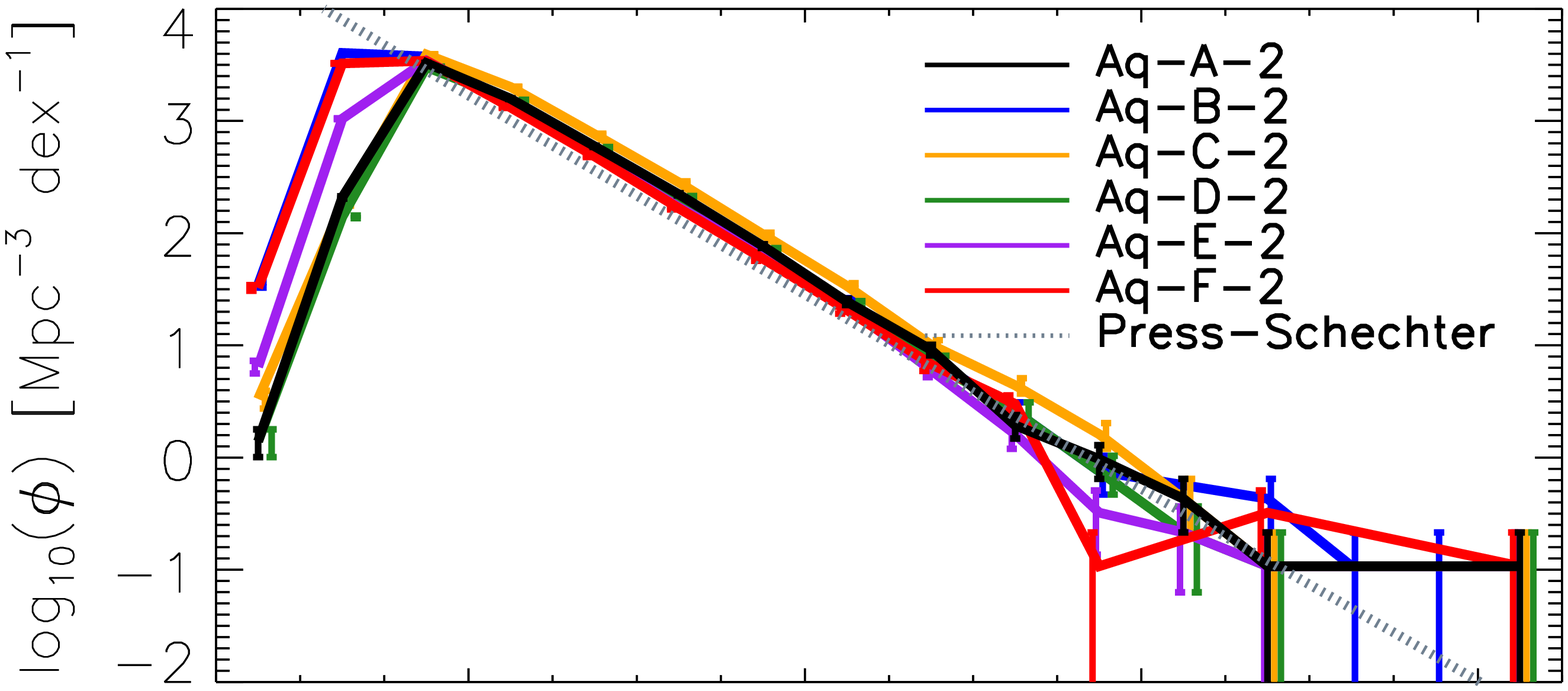}
  \label{fig:image1mv}
 \end{minipage}
 \hfill
 \begin{minipage}[t]{\linewidth}\vspace{-52pt}\hspace{-25pt}
     \includegraphics[width=1.2\linewidth]{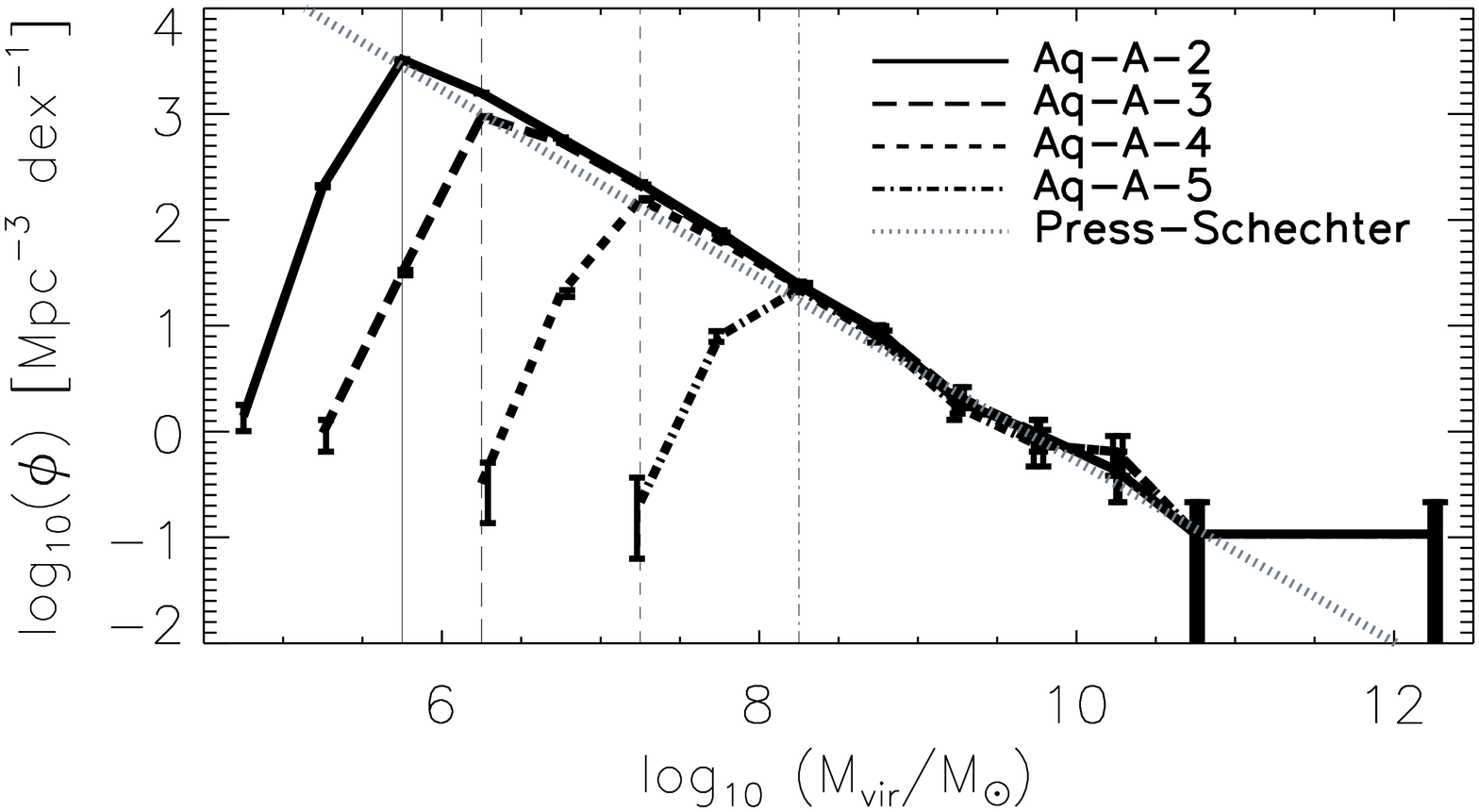}
   \label{fig:image2mv}
  \end{minipage}
  \caption{Upper panel: Virial mass function of central halos within
    \mbox{1.2 Mpc~$h^{-1}$} from the MW for all six Aquarius level--2
    simulations (coloured lines).  The dotted grey line corresponds to
    the fit to the mass function in our simulations using the publicly
    available code from \citet{Reed:2007} .  Lower panel: Virial mass
    function of the Aq-A halo for four different resolutions for
    central halos within \mbox{1.2 Mpc~$h^{-1}$} from the MW (black
    lines). The vertical lines show the different $M_{vir}$ threshold,
    according to each resolution: Aq-A-2 = 10$^{5.5}$ M$_{\odot}$,
    Aq-A-3 = 10$^{6}$ M$_{\odot}$, Aq-A-4 = 10$^{7}$ M$_{\odot}$,
    Aq-A-5 = 10$^{8}$ M$_{\odot}$.}
       \label{fig:mv}
\end{figure}

The upper panel of Figure \ref{fig:mv} shows the virial mass function
of central halos within 1.2 Mpc~$h^{-1}$ from the Milky Way--like
galaxies for the different Aquarius level--2 halos (coloured lines).
The lower panel shows the good convergence of the virial mass function
of the Aq-A halo in the four different resolutions.  The vertical
lines show the different $M_{vir}$ thresholds above which the mass
function has converged for each resolution level. The error bars in these plots, as well as similar
  figures in the rest of the paper, correspond to 1$\sigma$ uncertainties assuming 
  Poisson errors on the counts of each individual bin. In both panels the
grey dotted line corresponds to the \cite{Reed:2007} mass function fit
(using the Press--Schechter option in the code provided by these
authors).

\begin{figure}
 \begin{minipage}[t]{\linewidth}\vspace{0pt}\hspace{-25pt}
  \includegraphics[width=1.2\linewidth]{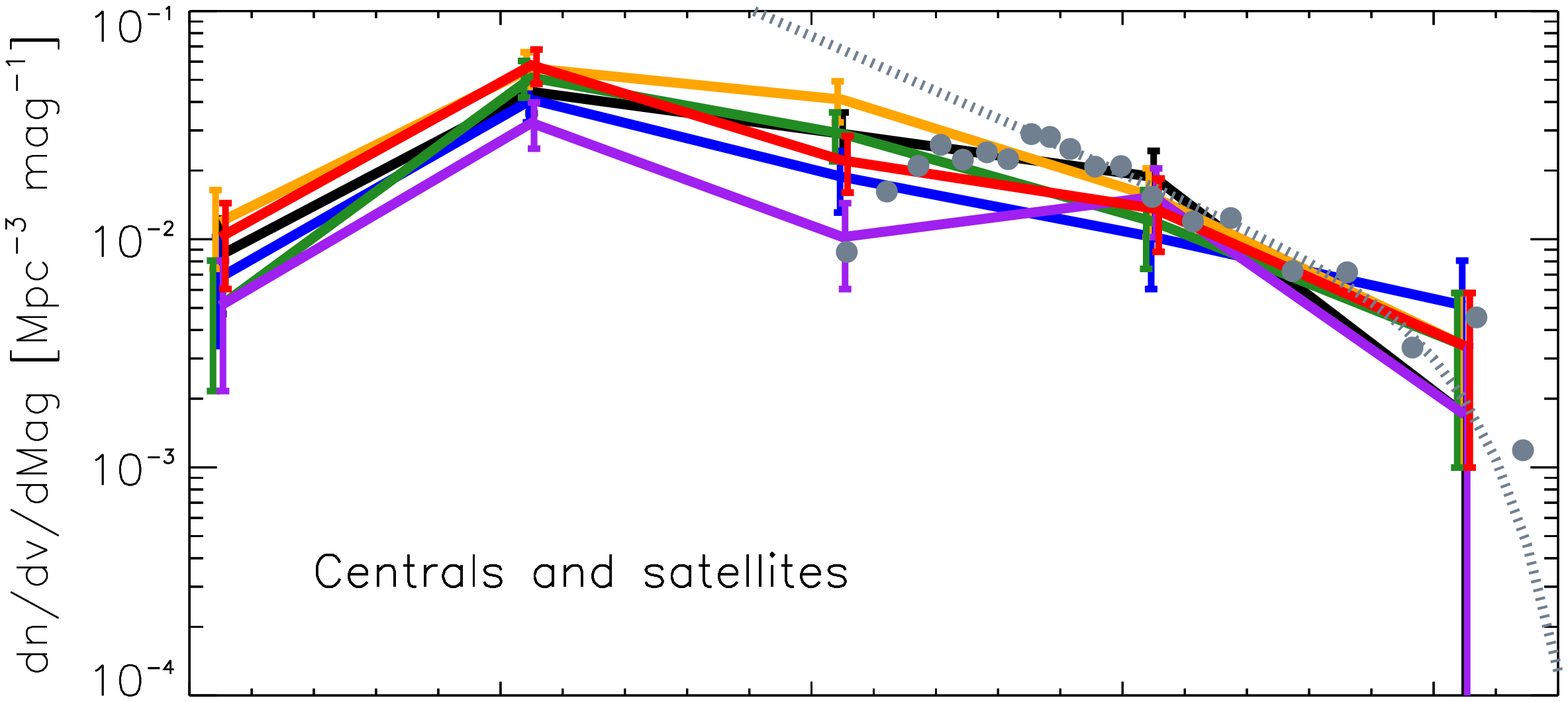}
  \label{fig:image1lf}
 \end{minipage}
 \hfill
\begin{minipage}[t]{\linewidth}\vspace{-52pt}\hspace{-25pt}
 \includegraphics[width=1.2\linewidth]{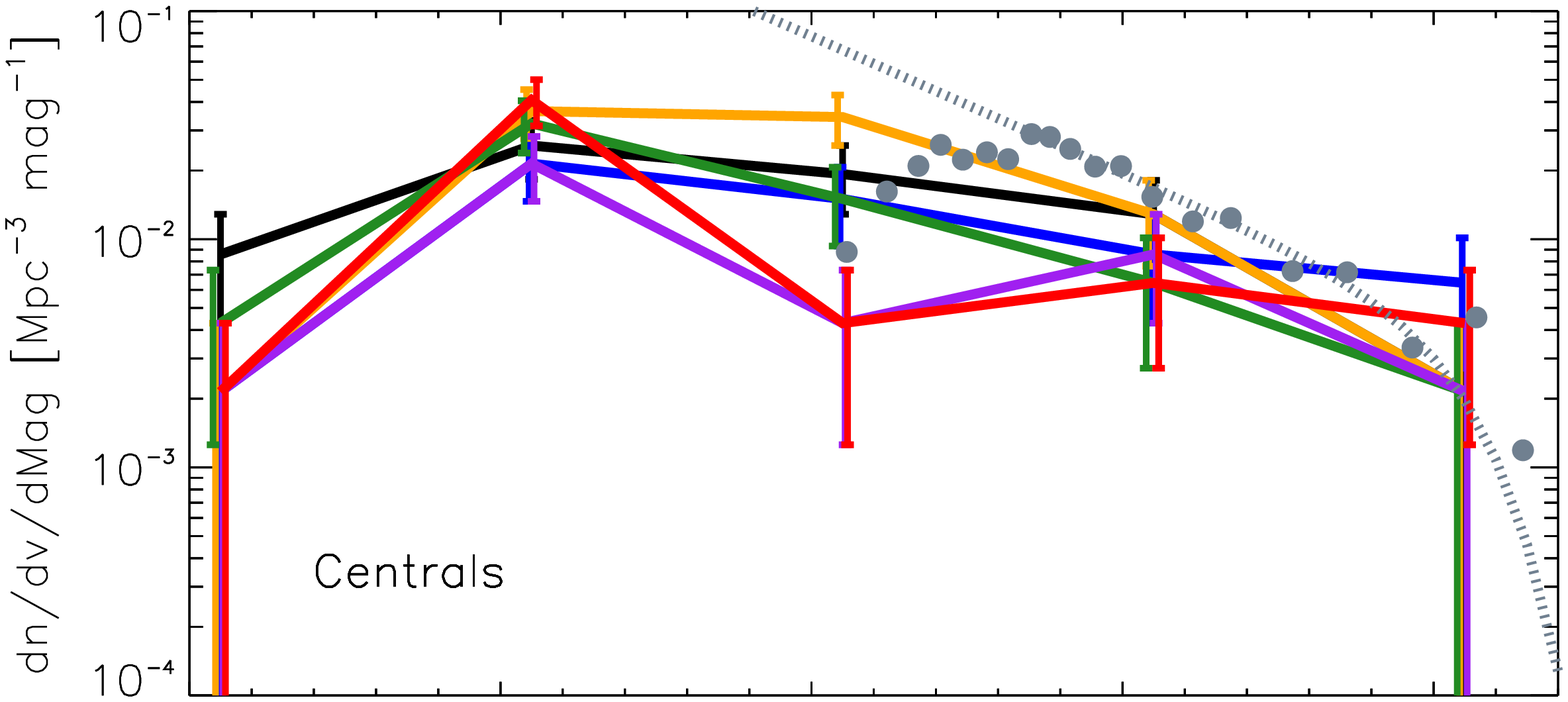}
  \label{fig:image2lf}
 \end{minipage}
 \hfill
\begin{minipage}[t]{\linewidth}\vspace{-52pt}\hspace{-25pt}
 \includegraphics[width=1.2\linewidth]{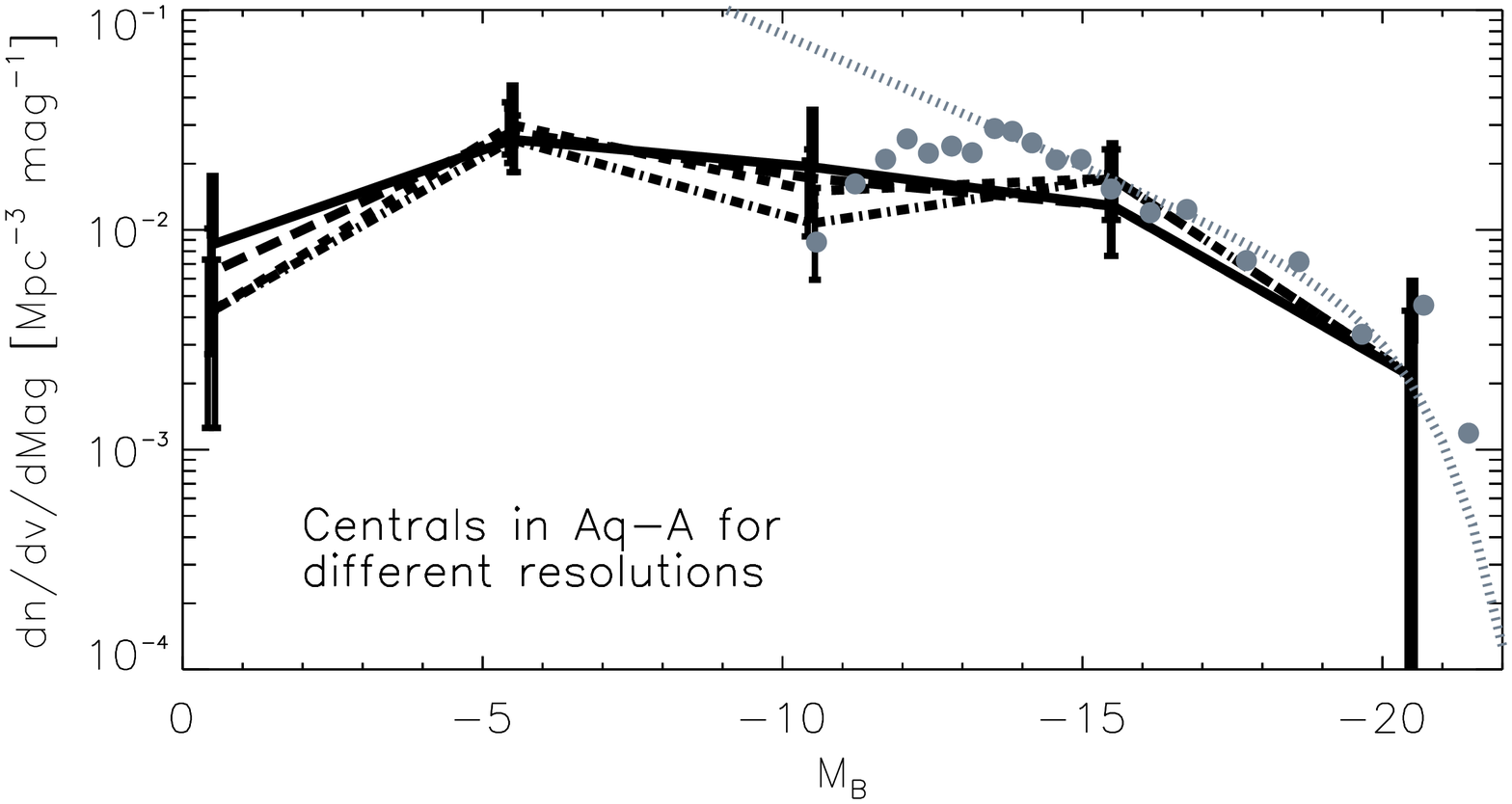}
 \label{fig:image3lf}
 \end{minipage}
 \caption{Upper panel: Luminosity function of central and
     satellites galaxies with \mbox{M$_{vir} > $ 10$^{5.5}$
       M$_{\odot}$} and within \mbox{1.2 Mpc~$h^{-1}$} from the main
     MW--like galaxy for all six halos in the Aquarius level--2 series
     (colours follow same scheme as in Fig.~\ref{fig:mv}).  The grey
     filled circles are from \citet{Klypin:2014} while the dotted grey
     curve shows the Schechter fit: $\phi^{*} = 1.25 \times
     10^{-2}$~Mpc$^{-3}$, $\alpha = 1.3$ and $M_{*,B} = -20.0 + 5
     \log(h)$. Middle and bottom panels: Luminosity function for
     central galaxies only. In the bottom panel only those of the Aq-A
     halo are shown for four different resolutions, and evidence the
   good convergence of the models. }
       \label{fig:lum}
\end{figure}

The luminosity function, i.e., the abundance of galaxies with a given
luminosity, is shown in Figure \ref{fig:lum} for our model. 
The top panel shows the luminosity functions including central and satellite
galaxies, while the bottom two are only for centrals. Different
colours correspond to galaxies residing in the different Aquarius
level--2 simulations.

To normalise these curves, and take into account the relatively small
volume sampled by the Aquarius simulations we computed $N_{Mill}$, the
mean number density of galaxies with B--band absolute magnitude in the
range $ -18 < M_B < -14$ in the milli--Millenium catalogue\footnote{
  http://gavo.mpa-garching.mpg.de/Millennium/} which uses the
semi--analytic model of \citet{Delucia2007}. In the same way we
computed $N_{AQ}$ for each of the Aquarius simulations, and estimated
the normalisation factor as $\log f = \log N_{AQ}/N_{Mill}$. 
Using only central galaxies, this is found to vary from 0.55 for
Aq-E to 0.85 for Aq-A, and this variation is simply due to small
number statistics. Therefore we have decide to take an average
normalisation factor $\log f = 0.7$ and applied this normalisation to
all our simulations. When using central and satellite galaxies still 
embedded in their own subhalo, the average normalisation factor is, 
as expected, only slightly different, with $\log f = 0.8$. 
Depending on the systems shown, these normalisations are also 
applied in Figures \ref{fig:HI} and \ref{fig:vel}. The black lines
in the bottom panel of Figure \ref{fig:lum} show how well the luminosity function has
converged by comparing the results for the four different resolutions
for central galaxies in the Aq-A series.

Many studies have estimated the luminosity function in the field
\citep[e.g.][]{Norberg:2002, Bell:2003, Blanton:2005,
  Trujillo-Gomez:2011}, however for the current study it is important
to know its shape at the faint end. This is why we compare here to the
luminosity function derived by \cite{Klypin:2014}.  These authors have
used the current version of Updated Nearby Catalog
\citep{Karachentsev:2013} which contains 869 galaxies with
redshift--independent distances \mbox{$D < 11$ Mpc} or radial
velocities with respect to centroid of the Local Group \mbox{$V_{LG} <
  600$ km $s^{-1}$}.  We focus on the subsample with distances
\mbox{$D \leqslant 10$ Mpc}, which comprises 733 galaxies, of which
652 objects are brighter than B--band absolute magnitude \mbox{$M_B = -10$} and 426 are brighter
than \mbox{$M_B = -13$}. The grey filled circles in Figure
\ref{fig:lum} represent the resulting luminosity function, while the
dotted grey curve presents a Schechter fit to this sample for galaxies
with \mbox{$M_B < -14$}, with the following Schechter parameters:
$\phi^{*} = 1.25 \times 10^{-2} h^3$ Mpc$^{-3}$, $\alpha = 1.3$ and
$M_{*,B} = -20.0 + 5 \log(h)$ \citep[Equation (2) of][]{Klypin:2014}.

This comparison shows that there is reasonable agreement between our
model and the field luminosity function for $M_B < -14$. The inclusion of satellites in the models does not
lead to a drastic change in the shape or normalisation of the luminosity function, especially if we consider the
important scatter from simulation to simulation. This is a consequence
of the small volume (and hence relatively small number of objects) of
the high--resolution region of the Aquarius halos. Coupled to possible
incompleteness in the data for galaxies with $M_B > -12$, and the
somewhat arbitrary normalisation, it is hard to argue that the
modelling needs any improvement.

\section{Results on the HI mass function and velocity function} \label{S_analysis}
\subsection{Neutral hydrogen (HI) mass function} \label{sS_HI}
\begin{figure}
 \begin{minipage}[t]{\linewidth}\vspace{0pt}\hspace{-25pt}
  \includegraphics[width=1.2\linewidth]{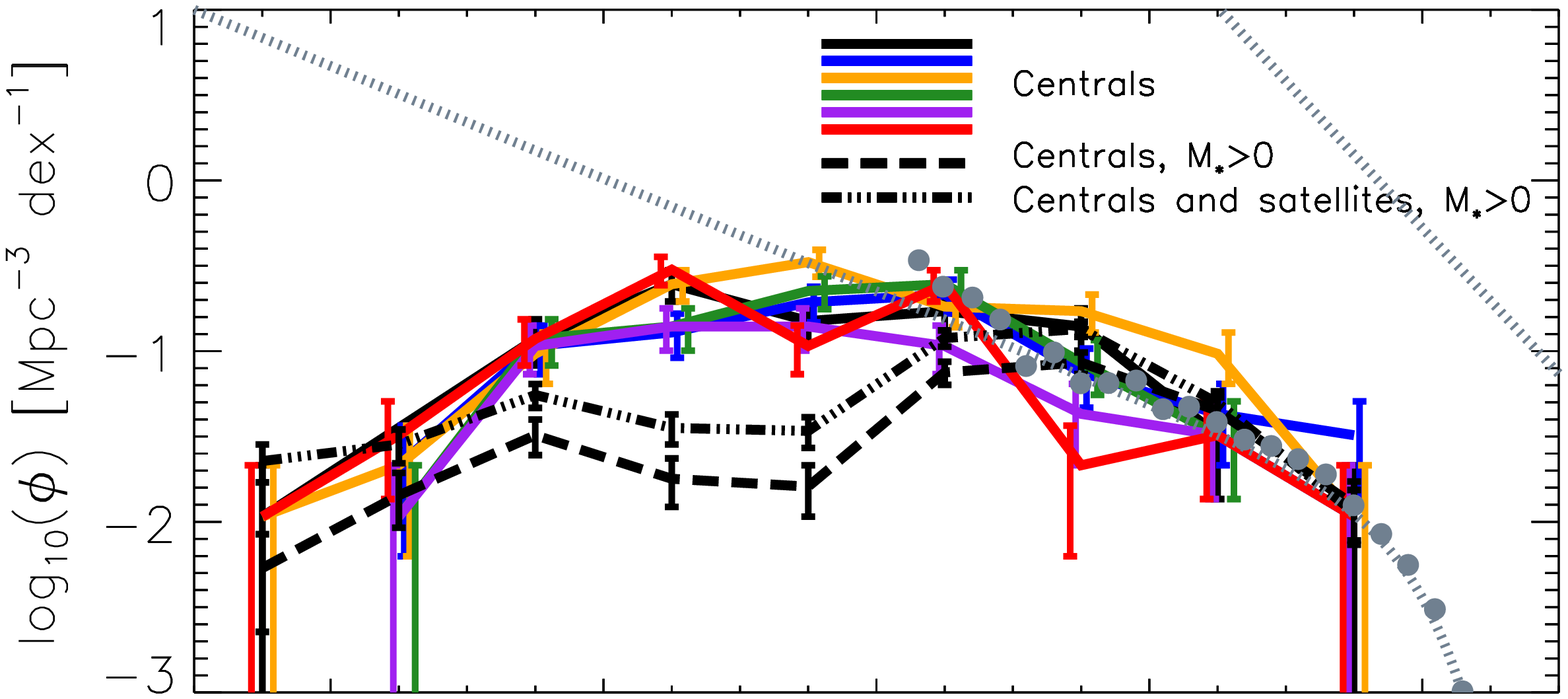}
    \label{fig:image1HI}
 \end{minipage}
 \hfill
 \begin{minipage}[t]{\linewidth}\vspace{-52pt}\hspace{-25pt}
  \includegraphics[width=1.2\linewidth]{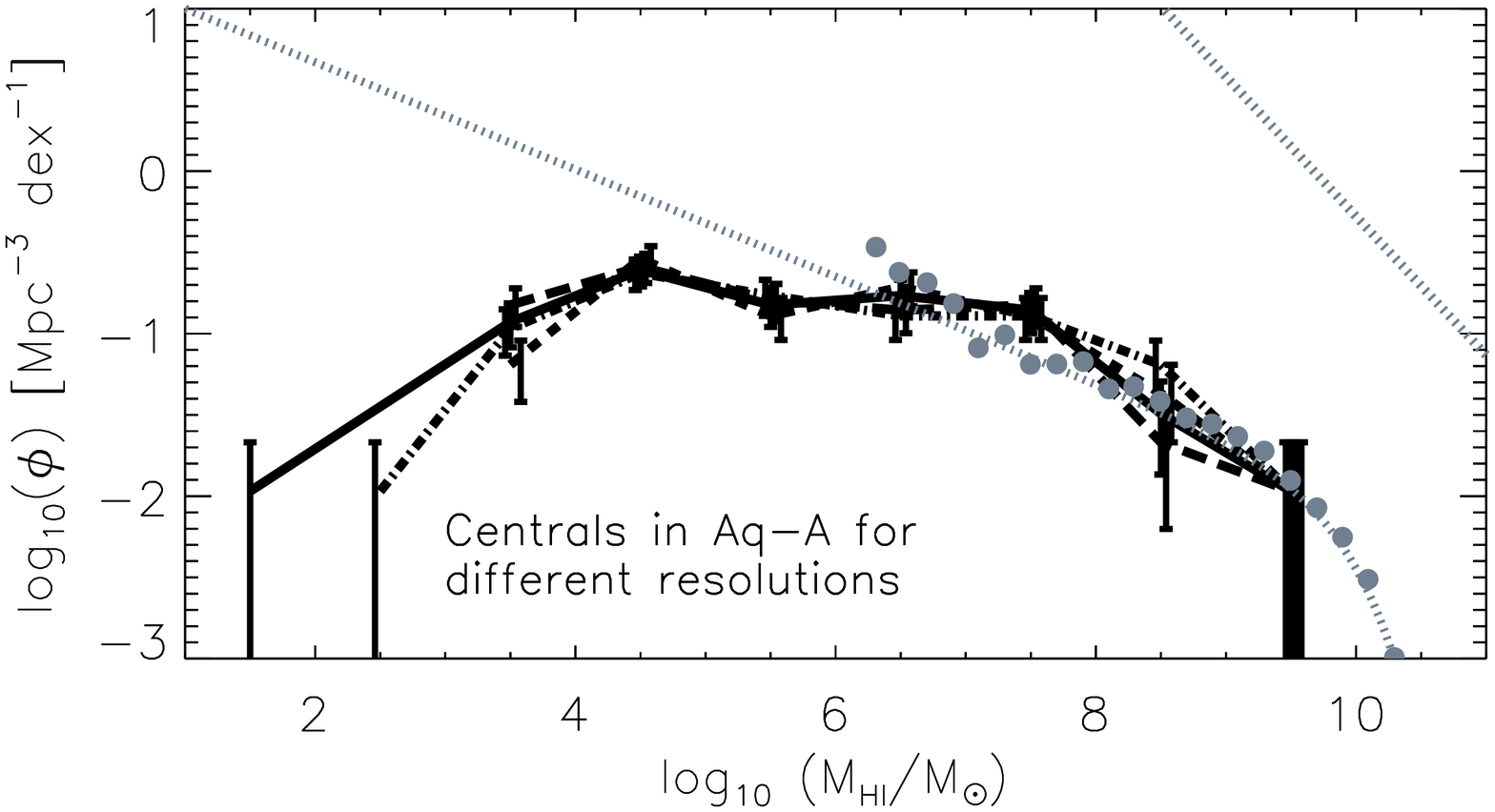}
   \label{fig:image2HI}
  \end{minipage}
  \caption{Upper panel: HI mass function of central galaxies for all
    six halos of Aq-2, with \mbox{M$_{vir} > $ 10$^{5.5}$ M$_{\odot}$}
    and within \mbox{1.2 Mpc~$h^{-1}$} from the MW (colour solid
    lines). The dashed curve shows the HI mass function for central
    galaxies containing both stars and gas, while the dot-dashed curve
    includes also satellite galaxies. Grey filled circles represent
    the HI mass function estimated while the grey dotted line is the
    Schechter function fit: $\phi^{*} = 0.0048$, $log(M_{*}) = 9.96$
    and $\alpha = -1.33$ \citep{Martin:2010}. The grey dotted line at
    the upper right--hand corner is the halo mass function.  Lower
    panel: HI mass function of the Aq-A halo for four different
    resolutions for central galaxies within \mbox{1.2 Mpc~$h^{-1}$}
    from the MW (black lines). We use different $M_{vir}$ thresholds
    for each resolution.}
       \label{fig:HI}
\end{figure}


Figures \ref{fig:mv} and \ref{fig:lum} confirm that the virial mass
function of halos and the luminosity function of galaxies have very different slopes at the low mass end. To
further explore this difference we now focus on the HI mass
function. This offers an independent test of the model because in low
mass systems the baryonic budget is dominated by gas rather than by
stars.

In the upper panel of Figure \ref{fig:HI} we show the HI mass
functions of central galaxies for all six Aquarius level--2 halos, for
which \mbox{M$_{vir} > $ 10$^{5.5}$ M$_{\odot}$} and located, as
before, in the high--resolution box around the main Milky Way--like
galaxies. Each colour corresponds to a different halo.  The
black--dashed line shows the HI mass function for central systems that
have {\it both} cold gas and stars, and is the average over all
Aquarius level--2 halos. It evidences that many objects with cold gas
(with $M_{HI} < 10^{6} \sm$) have not been able to form any stars (see
discussion below). The dot-dashed curve represents the 
corresponding HI mass function but now including both central and 
satellite galaxies, i.e. these are the counterparts of the objects 
shown in the top panel of Fig.~\ref{fig:lum}. Note that the two 
curves follow each other relatively well, and are only offset by 
$\sim 0.2$ dex at $M_{HI} \sim 10^{7.5} \sm$. This difference, due 
to satellite galaxies baring HI, is relatively small in comparison to 
the scatter from simulation to simulation. The black lines in the 
bottom panel of this figure show the HI mass function of the Aq-A halo
for the four different resolutions and again indicate very good convergence.

\cite{Martin:2010} have derived the HI mass function from a sample of
$\sim 10^4$ extragalactic sources comprising the ALFALFA $40\%$
survey (hereafter $\alpha.40$), with \mbox{$6.2 <
  \log(M_{HI}/\sm) < 11.0$}. These are plotted as the grey filled
circles in Figure \ref{fig:HI}.  The grey dotted line is the Schechter
function fit to this dataset with the parameters: $\phi^{*} = 0.0048$,
$\log(M_{*}) = 9.96$ and $\alpha = -1.33$ as estimated by
\citet{Martin:2010}.

The top panel of Figure \ref{fig:HI} shows that there is good 
agreement between the observed HI mass function of galaxies with 
$M_{HI} \sim 10^{6.5} - 10^{9} \sm$ and our models. There are, in 
fact, three comparisons to be made: $i$) with all central galaxies 
(including those without a stellar component, the coloured curves); 
$ii$) with central systems that host a luminous galaxy (dashed black 
curve); and $iii$) with centrals and satellites hosting a luminous 
galaxy (dot-dashed curve).  One of the reasons to consider these 
three classes separately rests on our aim to establish if the HI 
contents of central galaxies (in the field) are 
consistent with observations, in particular in relation to gauging 
the impact of environmental effects such as ram pressure stripping 
on satellites. Note that in all cases, at the high--mass end our 
simulation box is too small and does not contain enough galaxies, as 
evidenced by the large error bars in this figure.

The first point to note is the relatively large scatter from
simulation to simulation at intermediate masses for central galaxies,
although all curves are consistent within 1$\sigma$ of each other, and
on average also consistent with the observed HI mass
function. However, since most objects observed in the ALFALFA survey
actually have an optical counterpart \citep[see e.g.][]{Haynes2011}, a
better comparison to make is to the dashed black curve, representing
central luminous galaxies. We note the excellent agreement with the
observations at these masses, although for masses $M < 10^{6}\ \sm$,
the predicted HI mass function appears to decline to a point that may
be in tension with the observations. On the other hand, it is possible
that satellite galaxies are present in the observational sample. As
expected, when these are included in the predicted HI mass function,
there is an increase in the number of objects although the agreement
with the observations is still very reasonable. This implies that ram
pressure stripping, although it should be included when modelling the
properties of satellite galaxies, will not result in dramatic changes
that will break-down the good agreement between the models and the
observations, at least for the range of masses probed by the latter.

More careful comparisons, especially regarding the low mass end of the
HI mass function and the presence of systems without a stellar
counterpart, are necessary. It will be important to have both deeper
surveys in HI as well as in the optical, as these will allow us to
establish whether the trends predicted are correct.

\begin{figure}
 \begin{minipage}[t]{\linewidth}\vspace{0pt}\hspace{-25pt}
     \includegraphics[width=1.2\linewidth]{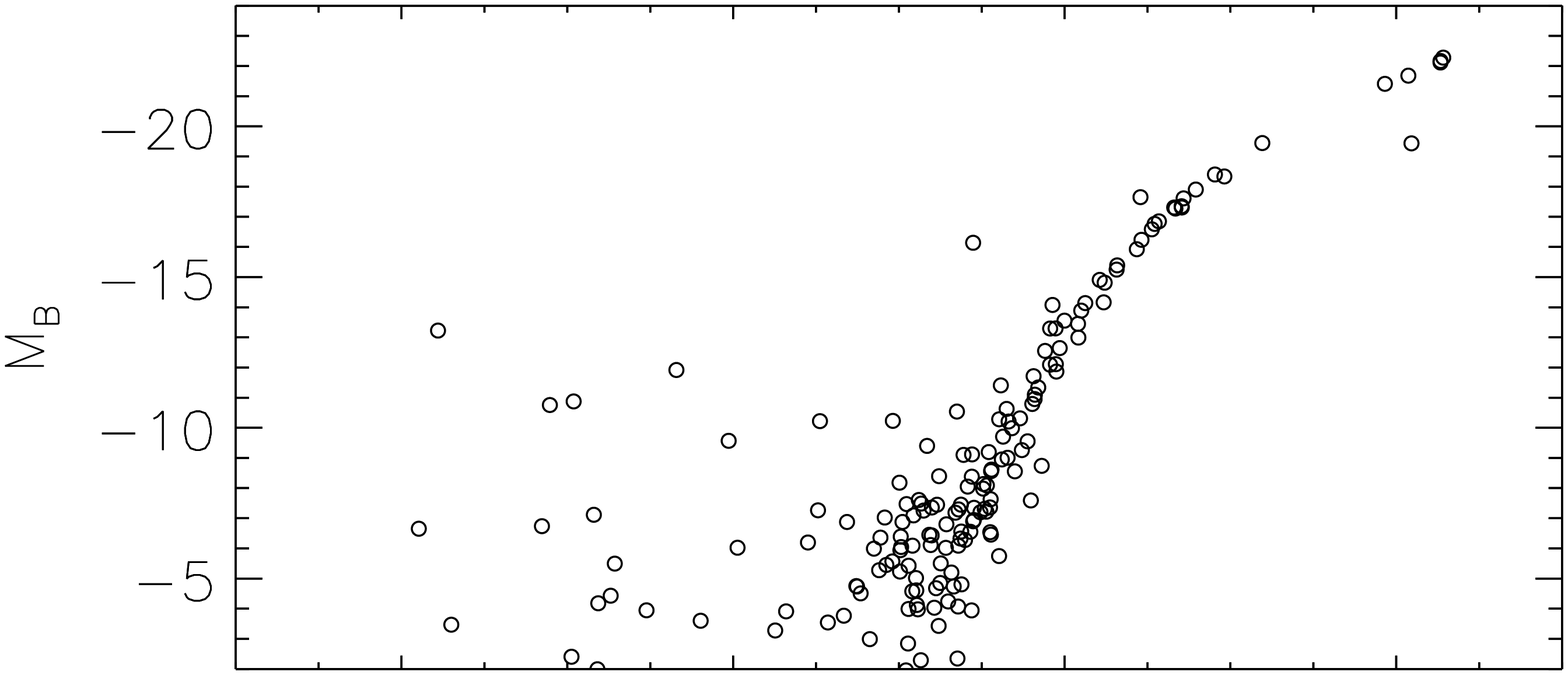}
   \label{fig:image1corr}
 \end{minipage}
\hfill
 \begin{minipage}[t]{\linewidth}\vspace{-52pt}\hspace{-25pt}
      \includegraphics[width=1.2\linewidth]{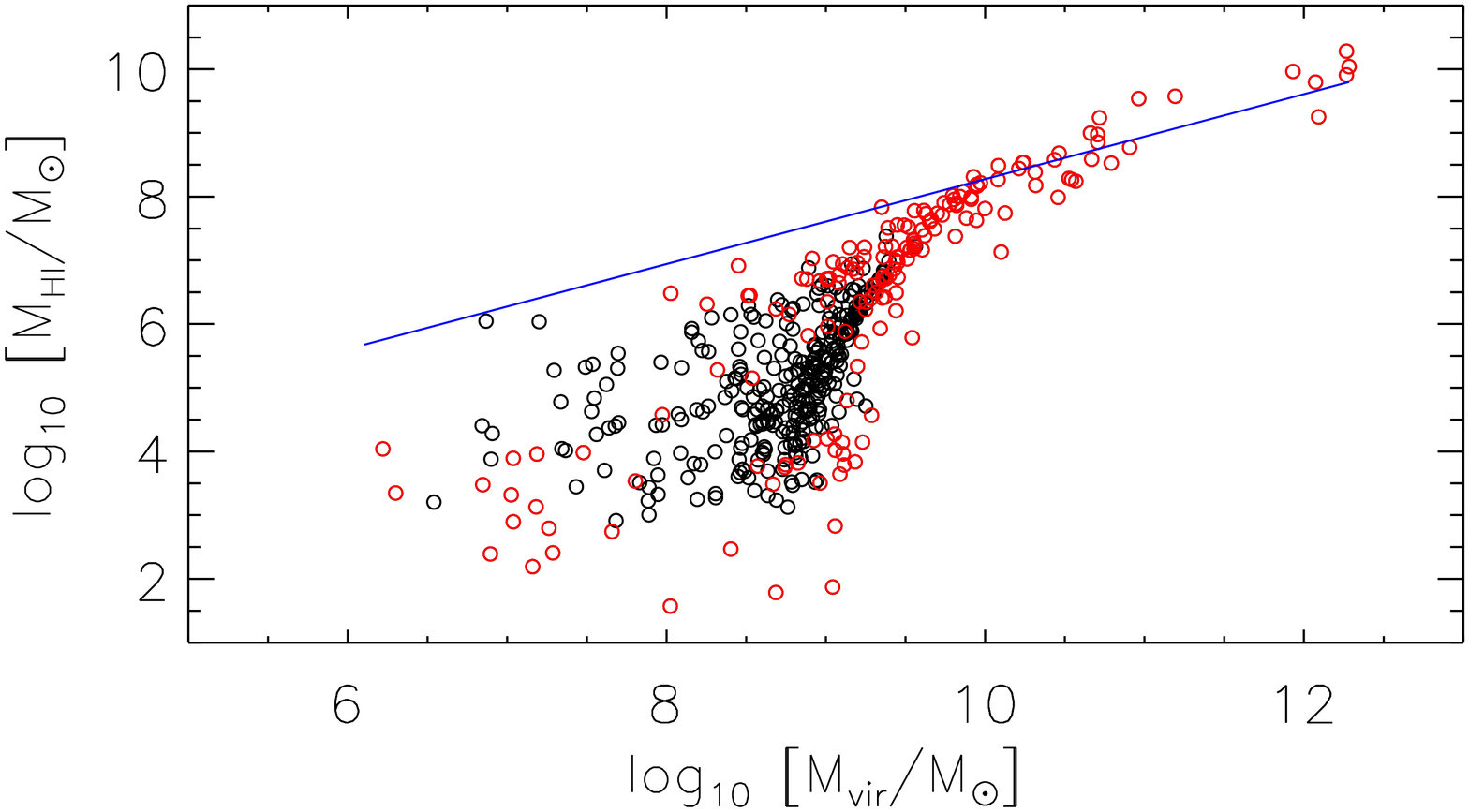}
      \label{fig:image2corr}
  \end{minipage}
  \caption{Upper panel: $M_B$ vs $M_{vir}$ for all the central galaxies in our
    models.  Lower panel: $M_{HI}$ vs $M_{vir}$ for objects with stars
    (red circles) and for those with only gas (black circles). The
    line shows $\tilde{M}_{crit}$, an upper limit estimate of the critical mass 
    above which a halo would form stars}. 
       \label{fig:props}
\end{figure}

In the top panel of Figure \ref{fig:props} we have plotted the B--band absolute magnitude 
against the virial mass for all the 173 central galaxies in the
high--resolution region of the Aquarius halos. The panel below shows
the HI mass against the virial mass for all these galaxies as red
circles, while the open circles denote objects that according to our
SA model have not formed any stars ($M_* = 0$) but do contain cold
gas. There are 315 such objects with $M_{vir} \ge 10^{5.5} \sm$. The
reason such systems do not form stars is because their surface gas
density is below the threshold for star formation imposed by our
model. 

An estimate of the dependence of the critical mass on the virial mass
of objects can be obtained using Eq.~(\ref{eq:mcrit}). In that
equation, we replace the disk radial scalelength by $\tilde{r}_d =
\lambda r_{vir}/\sqrt{2} $ where $\lambda$ is the dimensionless spin
parameter (which we take to be $\sim 0.035$) and $r_{vir}$ the virial
radius of the host halo. As explained earlier, this scale is obtained assuming conservation
of specific angular momentum, and that when hot gas cools at the
centre of dark matter haloes, it settles in a rotationally supported
disk \citep{Mo:1998}. The value of $\tilde{r}_d$ will typically be
larger than $r_d$, because the latter is a mass--weighted average over
the whole gas cooling history of the system.  Therefore
$\tilde{M}_{crit}$ obtained in this way will be overestimated in
general. Nonetheless, this rough zero order approximation of the
critical mass is plotted as the line in the bottom panel. This
comparison serves to tentatively support our claim that the density of
cold gas in lower mass objects is too low to allow star formation
taking place. 

It should also be borne in mind that since the criterion given by Eq.~(\ref{eq:mcrit}) 
for the critical mass of gas for star formation is 
strict and global, it is likely that if some stochasticity
would be allowed and density variations inside a system considered,
that some systems below the threshold should be able to form
stars. This implies that the exact predictions for the shape of the HI mass
function for objects around and below $M_{crit}$ should be taken with
a grain of salt.

This analysis supports the idea that ram pressure 
stripping of gas once a system becomes a satellite will not affect 
the properties of the objects significantly. Most of the gas present 
in the systems has too low density to form stars, which is 
also why the luminosity function of galaxies (including satellites) 
is in agreement with observations, even if in our model, 
ram pressure has not really been implemented. For this reason, 
in the remainder of this paper we only focus on the behaviour of central galaxies.

\subsection{Velocity function} \label{sS_Vel}

So far we have compared the observed global baryonic properties of
galaxies to those in our models. An additional important complementary
aspect concerns the internal dynamics of galaxies, which links the
galaxies to their host dark matter halos. A fundamental question that
potentially relates to the nature of dark matter, is whether the
galaxies in the SA model are hosted by the right dark matter halos in
the simulations.
 
This is why we now focus on velocity function, namely the abundance of
galaxies with a given circular velocity \citep{Cole:1989,
  Shimasaku:1993, Gonzalez:2000, Zavala:2009, Trujillo-Gomez:2011}. To
measure circular velocities for large numbers of galaxies is
challenging, but wide--area, single--dish 21~cm surveys are making this
possible. Although ideally one would like to obtain full rotation
curves for a large number of systems spanning a large range in galaxy
mass, this is too time--consuming to be currently feasible. Therefore,
what is rather used is the width of the spectral HI line profile
$w$. This is believed to be close to the peak rotational velocity of
the system at approximately two radial scale--lengths.

\cite{Papastergis:2011} have measured the velocity width function (WF) of
HI--bearing galaxies down to $ w = 20~\kms$ in the ALFALFA survey.  This WF is based on 10,744
HI--selected galaxies from the $\alpha.40$ survey (a more than twofold
increase over previous data sets), so it is the largest HI--selected
sample to date.

For each dark matter halo in the Aquarius suite \citet{Springel:2008a}
has derived the peak circular velocity $V_{max}$ and the circular
velocity at the virial radius $v_c(r_{vir}) = V_{vir}$, as well as the
location of the peak $r_{max}$ and the virial radius $r_{vir}$. With
this information, and assuming an NFW shape, we may derive the full
form of the rotation curve. Furthermore, we may derive $v_{rot}$,
i.e. $v_c(2 r_d)$ using the value of $r_d$ as determined from our SA
model \footnote{Our results do not change significantly if we instead
  compute the circular velocity at $r_d$ or at $3 r_d$.}

To compare this data with the results of \cite{Papastergis:2011}, we convert rotational 
velocities into HI velocity widths by assuming the relationship given by Equation (4) 
of \cite{Papastergis:2011}: 
\begin{equation}
\mbox{$w = 2v_{rot} \sin i + w_{eff}$} 
\end{equation}
This equation indicates that the galaxies are randomly oriented with
respect to the line of sight ($\cos i$ is uniformly distributed in the
[0, 1] interval), while $w_{eff}$ is a small "effective" term used to
reproduce the broadening effect of turbulence and non--circular motions
on HI linewidths.  Following \cite{Papastergis:2011} we consider
\mbox{$w_{eff} = 5~\kms$} for the broadening term, which is added
linearly for galaxies with \mbox{$v_{rot} > 50~\kms$} and in quadrature
for lower velocity galaxies.

\begin{figure}
 \begin{minipage}[t]{\linewidth}\vspace{0pt}\hspace{-25pt}
  \includegraphics[width=1.2\linewidth]{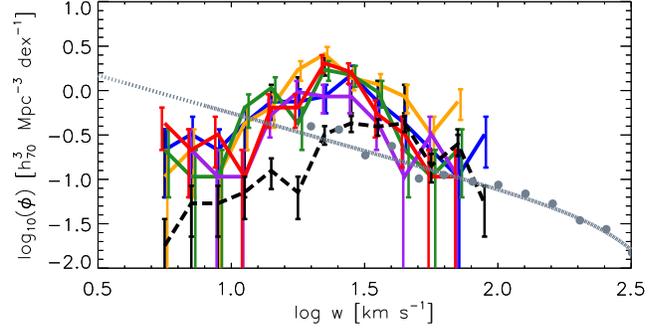}
\end{minipage}
 \caption{Velocity width function of central galaxies within \mbox{1.2
     Mpc~$h^{-1}$} from the centre of the MW--like Aquarius halos, with
   \mbox{M$_{HI} > $ 10$^{4}$ M$_{\odot}$} and with \mbox{M$_{vir} > $
     10$^{5.5}$ M$_{\odot}$} for the all six halos of resolution
   level--2 (coloured lines). The dashed curve is the average over all size
   Aquarius halos taking into account only those systems hosting also stars. Grey
   filled circles represent the measured ALFALFA WF, and dotted grey
   curve represents the fit of the modified Schechter function:
   $\phi^{*} = 0.011 h_{70}^{3} Mpc^{-3}$, $\log w_{*} = 2.58$, $\alpha = -0.85$ and $\beta =2.7$
   \citep{Papastergis:2011}.}
       \label{fig:vel}
\end{figure}

Figure \ref{fig:vel} shows the resulting velocity width function of
the same central galaxies as in previous plots, with \mbox{$M_{HI} > $
  10$^{4}~\sm$} and with \mbox{$M_{vir} > $ 10$^{5.5}~\sm$} for the all six halos of resolution level--2 in the
Aquarius suite (coloured lines).  The velocity function considering only systems with a stellar
counterpart (i.e. $M_* > 0$) averaged over all six Aquarius halos is
shown as the dashed black curve.

The grey filled circles in the same figure represent the measured
ALFALFA WF presented by \cite{Papastergis:2011}, based on galaxies
which are positioned in the portion of the flux--width plane where the
ALFALFA survey is complete and have profile widths broader than
\mbox{$w \gtrsim 20$ km $s^{-1} $}.  The dotted grey curve represents
the fit of the modified Schechter function \citep[Equation (2)
of][]{Papastergis:2011} with the following parameters: $\phi_{*} =
0.011 h_{70}^{3}$ Mpc$^{-3}$ dex$^{-1}$, log $w_{*} = 2.58$, $\alpha = -0.85$, and $\beta =2.7$
\citep{Papastergis:2011}.

The agreement between the predictions of the model and the
observations is quite good. It shows that our model places
galaxies of the right baryonic content (in stars and in HI gas) in the
``right'' dark matter halos. Contrary to previous work \citep[see
e.g.][ and references therein]{Papastergis:2014}, our model does not
present an excess of luminous galaxies with too low velocity widths
compared to the observations. The resulting function has a similar
slope as observed. 

It is interesting to note that in our model, those systems with cold
gas but lacking stars, lead to an excess above the Schechter function
that fits so well the luminous galaxies in our model. This excess only
appears for $\log w \sim 1.2 - 1.5$, which is right around, and slightly
below the limit of the ALFALFA survey.

\subsection{Zooming into the properties of galaxies and their host halos}

Our models do not appear to have a significant excess
of halos hosting galaxies of a given velocity width $w$ at the
faint/low mass end. To understand further this result we now zoom into the
properties of the halos hosting galaxies.  

\citet{Papastergis:2014}
analysed the kinematics of a sample of gas--rich dwarf galaxies
extracted from the literature. They derived $v_{rot}$ by making
inclination corrections using SDSS images, and estimated the value of 
$V_{max}$ by considering the most massive halo that is consistent with
the last measured point of the rotation curves available for these
galaxies.

\begin{figure}
 \begin{minipage}[t]{\linewidth}\vspace{0pt}\hspace{-25pt}
  \includegraphics[width=1.2\linewidth]{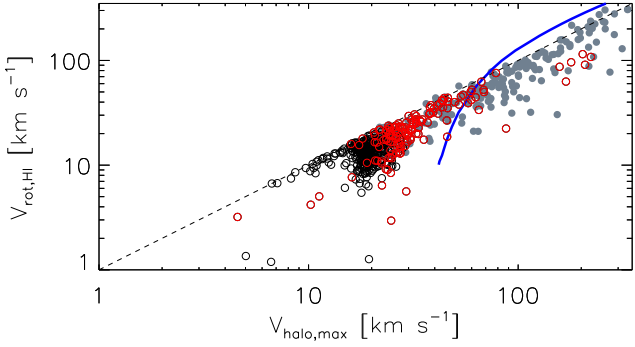}
   \label{fig:image1}
 \end{minipage}
 \hfill
 \caption{Distribution of $v_{rot}$ vs $V_{max}$ for the dataset from
   \citet{Papastergis:2014} (grey solid points), for the dark matter
   halos hosting our model galaxies (red circles), and for systems in
   our model that have $M_* = 0$ but $M_{HI} > 10^4 \sm$ (black circles). The blue
   curve shows the relation predicted by abundance matching given by  \citet{Papastergis:2014}. The
   luminous central galaxies in our model occupy the same region as the
   observations. Note that for $V_{max} > 100~\kms$, the rotational 
   velocities of the observed galaxies are systematically higher. This
   merely reflects that the baryon contribution (which is not taken
   into account for the model galaxies), is non--negligible for these
   systems.}
       \label{fig:vmax_vrot}
\end{figure}

In Figure ~\ref{fig:vmax_vrot} we show the distribution of $v_{rot}$
vs $V_{max}$ for this dataset (grey solid points) as well as for th 
central dark matter halos hosting our model galaxies (red circles). With black
open circles with indicate those systems in our model that have $M_* =
0$ but $M_{HI} > 10^4 \sm$, i.e. they are devoid of stars, and hence
would not be in the observational compilation of
\citet{Papastergis:2014}. The blue curve shows the relation predicted
by abundance matching according to \citet{Papastergis:2014}.

This comparison is extremely satisfactory and confirms again that
model galaxies are placed in the right host dark matter halos. No
discrepancies are found, and there is no evidence from this plot that
the dark matter halos in the simulations would be too massive or too
dense to host the observed galaxies. 

However, this is in tension with the naive conclusion that one would
have drawn from comparing the observations to the predictions of
abundance matching. It is important therefore to understand why the
galaxies in our model do not follow the predictions of abundance
matching. Equally important it is to understand why only halos with
$V_{max} > 20~\kms$ host luminous galaxies.

The reason that abundance matching does not work at the low mass end
of the halo spectrum is that the ability of a halo to host a galaxy
depends on its capacity to i) retain the baryons; ii) cool gas from the hot phase; and
iii) have enough cold gas at high density for star formation. These
conditions are redshift dependent, and hence the virial mass or
velocity at the present--day are not sufficient to establish whether a
halo will host a galaxy (unless this mass is significantly different from the
various thresholds). Of the halos that are near the thresholds for these
physical processes to take place, some fraction will satisfy these
requirement, and this fraction is not fully random, and is not mass
ranked (as in abundance matching) but depends on the specific history of the
halo. 

\begin{figure}
 \begin{minipage}[t]{\linewidth}\vspace{0pt}
  \includegraphics[width=1.2\linewidth]{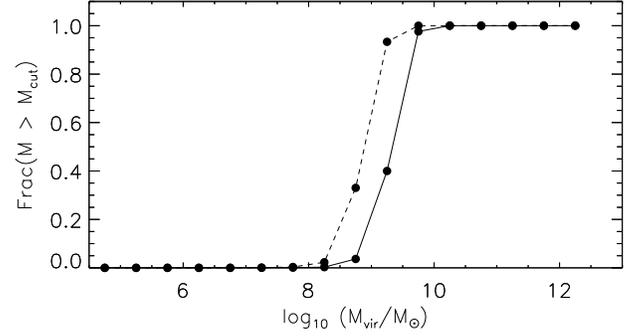}
\end{minipage}
 \hfill
 \caption{Fraction of halos hosting a luminous galaxy with $M_* > 10^3 \sm$ (solid) 
         and halos with more than $10^4\sm$ HI gas (dashed) as function of $M_{vir}$.}
       \label{fig:mass-fractions}
\end{figure}

This point is made clear by considering the fraction of central halos that
host galaxies as a function of virial mass. This is plotted in Figure
\ref{fig:mass-fractions}, where the fraction of galaxies with $M_* >
10^3 \sm$ at a given $M_{vir}$ is indicated by the solid line, and the
fraction of systems with $M_{HI} > 10^4 \sm$ is given by the dashed
line. From this figure we see that all halos with $M_{vir} > 10^{10}
\sm$ host luminous galaxies, while only $50\%$ of those with $M_{vir}
\sim 10^{9.4} \sm$ do. On the other hand, this function is shifted to lower masses if we
consider the systems with cold gas: 50\% of the halos with
$M_{vir} = 10^{8.9} \sm$ contain HI. This
shift is present because, not only must a halo be able to cool gas but
in order to form stars, this gas must be above the threshold for star
formation. It is also interesting that these functions are not exactly steps,
and this is a manifestation of the various processes at work which have different 
thresholds, as discussed above.

\begin{figure}
 \begin{minipage}[t]{\linewidth}\vspace{0pt}\hspace{-25pt}
  \includegraphics[width=1.2\linewidth]{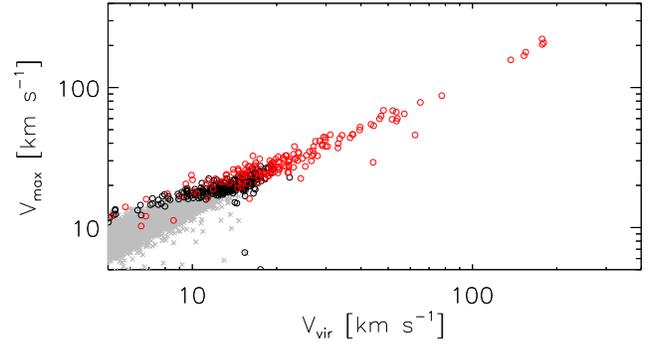}
\end{minipage}
 \hfill
 \caption{Distribution of $V_{vir}$ vs $V_{max}$ for the central dark matter
   halos hosting our model galaxies (red circles), for systems in our
   model that have $M_* = 0$ but $M_{HI} > 10^4 \sm$ (black open
   circles) and for completely dark halos (grey crosses). }
       \label{fig:vmax_vvir}
\end{figure}

Figure \ref{fig:vmax_vvir} shows the $V_{max}$ vs $V_{vir}$ for the central model galaxies as
red circles, the systems with no stars but with $M_{HI} > 10^4 \sm$ as
open black circles, and the dark matter halos without baryons as
grey crosses. This demonstrates clearly that virial velocity today
$V_{vir}$ is not enough to predict if a halo will host a luminous
galaxy. It also shows that at a given $V_{vir}$, luminous galaxies are
hosted in halos with higher $V_{max}$. This implies that for a given
$V_{vir}$ these are the most concentrated halos since for a NFW halo:
\begin{equation}
V_{max} = 0.465 V_{vir} \left(\frac{c}{f(c)}\right)^{1/2}
\label{eq:vmax-vvir}
\end{equation}
and 
\begin{equation}
f(c) = \ln (1+c) - c/(1+c)
\label{eq:fc}
\end{equation}
where $c = r_{vir}/r_s$ is the concentration, with $r_s$ the scale radius of the halo.

These more concentrated halos will have typically collapsed earlier,
implying that they reach a higher mass at earlier times which
therefore increases their likelihood of being above e.g. the HI
cooling limit at which point they can start forming stars if they have
retained enough baryons (after photoevaporation).

This also explains why the hosts of luminous galaxies are to the left of 
the abundance matching curves, since
\begin{equation}
\frac{V^2_{rot}}{V^2_{max}} = 4.63 \frac{f(x)}{x} 
\end{equation}
where $f(x)$ is given by Eq.~(\ref{eq:fc}) with $x = 2 r_d/r_s$. In
general, we can assume that $x \lesssim 2$ (i.e. $r_d < r_s$), in
which case $f(x)/x$, increases with increasing $x$. This implies that
for fixed $r_d$, galaxies with smaller $r_s$ (i.e. in more
concentrated halos), have larger $V_{rot}/V_{max}$. In other words
$V_{max}$ at fixed $V_{rot}$ is lower for more concentrated halos.
This highlights that for abundance matching to work, it is important
to select halos that are most concentrated
amongst those that are near the thresholds. 

Figure \ref{fig:vmax_vvir} shows that dark halos (grey crosses) have
$V_{vir} < 20 ~\kms$, which is consistent with the fact that the HI cooling
limit is $V_{vir}^{cool} \sim 16.7~\kms$. Eq.~(\ref{eq:vmax-vvir})
explains why basically all objects with $V_{max} \lesssim 20~\kms$ are
dark, as the function $\sqrt{c/f(c)}$ is weakly dependent on $c$. Since
halos hosting luminous galaxies are on average in more concentrated
halos as discussed above, their $V_{max}$ is higher as seen in
this figure.

\section{Conclusions} \label{S_Conclusions}

We have used the high--resolution Aquarius cosmological dark matter
simulations in combination with the semi--analytic model by
\citet{Starkenburg:2013} to study the HI content and dynamical
properties of galaxies at the low mass end in the context of the
\mbox{$\Lambda$CDM} paradigm. We have compared our predictions to the
observed ALFALFA survey, and found excellent agreement with the HI
mass and velocity width functions measured by this survey for 
central galaxies, down to the lowest mass scales probed. 
Implicit in this conclusion, is our assumption that luminous 
satellite galaxies do not contribute significantly to these HI 
distribution functions. This is based on the fact that even if these 
systems are considered when computing the model HI mass function, 
we still find good agreement with observations, and the same is 
true for the luminosity function. Therefore, even if we do not 
include the effect of ram pressure stripping in our models 
explicitly, we argue that its effect will not lead to dramatic 
changes in the properties of galaxies. This is because the gas that is 
present in these systems is too diffuse to contribute to star 
formation, and hence have an impact on the evolution of these objects.

We have also studied the relation between two global parameters of 
the circular velocity curves of the galaxies in our models and in 
the ALFALFA survey, namely between their peak velocity and circular 
velocity at 2 radial disk scalelengths. The distribution found in 
our models overlaps perfectly with that inferred from observations. 
This suggests that our galaxies are placed in the right dark matter 
halos, and consequently and at face value we do not seem to find any 
discrepancy with the predictions from the \mbox{$\Lambda$CDM} model.

Our model predicts the existence of a population of HI halos that do
not have a stellar counterpart \citep[e.g.][]{Salvadori:2012}. 
The exact abundance of these objects is likely to depend rather strongly on 
the implementation of star formation in the models, including whether 
cooling below $10^4 K$ is allowed (via $H_2$) and on the critical density 
floor for star formation. Although these objects are relatively gas--rich, 
their cold gas is too diffuse. Furthermore, their baryon fraction is well below
universal, since typically $f_b \le 0.01$, reflecting the fact
that they lost an important fraction of baryons because of
photoevaporation during reionization.
\\

\label{acknowledgements}
We are grateful to Manolis Papastergis and Simon White for interesting
discussions and the Aquarius project members, especially to Volker
Springel.  We are also indebted to Gabriella De Lucia and Yang-Shyang
Li for the numerous contributions in the development of the
semi-analytic code used here. The referee, Darren Croton, is
acknowledged for a very positive and constructive report that has helped improve the
manuscript.  This work has been partially supported by the Consejo de
Investigaciones Cient\'ificas y T\'ecnicas de la Rep\'ublica Argentina
(CONICET), by the Secretar\'ia de Ciencia y T\'ecnica de la
Universidad Nacional de C\'ordoba (SeCyT) and by the European
Commissions Framework Programme 7, through the International Research
Staff Ex--change Scheme LACEGAL.  AH acknowledges financial support
from the European Research Council under ERC--StG grant
GALACTICA--240271. AH and MA acknowledge grant PICT1137 from FONCYT
Argentina. ES acknowledges partial funding from the Canadian Istitute
for Advanced Research (CIFAR).

\label{lastpage}

\footnotesize{
  \bibliographystyle{mn2e}
  \bibliography{Bibliography.bib}
}

\end{document}